\documentclass[%
onecolumn,%
oneside,%
floats,%
aps,%
prd,%
nobibnotes,%
nofootinbib,%
amsmath,%
amssymb,%
amsfonts,%
amscd,%
superscriptaddress,%
eqsecnum%
]{revtex4}

\usepackage[utf8]{inputenc}
\usepackage{graphicx,array,dcolumn,amsmath,amssymb}
\usepackage{cases}
\usepackage[paperwidth=210mm,paperheight=297mm,centering,hmargin=1.8cm,vmargin=2.5cm]{geometry}

\newcommand{\bi}{\bibitem}
\newcommand{\be}{\begin{equation}}
\newcommand{\ee}{\end{equation}}

\def\b{\textbf}

\def\be{\begin{eqnarray}}
\def\ee{\end{eqnarray}}

\def\-g{\sqrt{-g}}

\renewcommand\rho{\varrho}
\renewcommand\tilde{\widetilde}

\begin{document}

\title{Jeans instability in classical and modified gravity}

\author{E.V. Arbuzova}
\email{arbuzova@uni-dubna.ru}
\affiliation{Novosibirsk State University, Novosibirsk, 630090, Russia}
\affiliation{Department of Higher Mathematics, University ``Dubna'', 141980 Dubna, Russia}

\author{A.D. Dolgov}
\email{dolgov@fe.infn.it}
\affiliation{Novosibirsk State University, Novosibirsk, 630090, Russia}
\affiliation{ITEP, Bol. Cheremushkinsaya ul., 25, 113259 Moscow, Russia}
\affiliation{Dipartimento di Fisica e Scienze della Terra, Universit\`a degli Studi di Ferrara\\
Polo Scientifico e Tecnologico - Edificio C, Via Saragat 1, 44122 Ferrara, Italy}

\author{L. Reverberi}
\email{reverberi@fe.infn.it}
\affiliation{Dipartimento di Fisica e Scienze della Terra, Universit\`a degli Studi di Ferrara\\
Polo Scientifico e Tecnologico - Edificio C, Via Saragat 1, 44122 Ferrara, Italy}
\affiliation{Istituto Nazionale di Fisica Nucleare (INFN), Sezione di Ferrara\\
Polo Scientifico e Tecnologico - Edificio C, Via Saragat 1, 44122 Ferrara, Italy}

\begin{abstract}
Gravitational instability in classical Jeans theory, General Relativity, and modified gravity is considered. The background density increase leads to a faster growth of perturbations in comparison with the standard theory. The transition to the Newtonian gauge in the case of coordinate dependent background metric functions is studied. For modified gravity a new high frequency stable solution is found.
\end{abstract}

\maketitle

\section{Introduction \label{s-introduction}}

The instability of self-gravitating systems was first investigated by Jeans~\cite{Jeans} in non-relativistic Newtonian gravity. It was extended to General Relativity (GR) by Lifshitz~\cite{Lifshitz} and nowadays it is widely used in cosmology to study the rise of density perturbations in the expanding universe~\cite{ZN,Mukha,SW,GR}. The original Jeans approach is based on the Poisson equation 
\be
\Delta \Phi =4\pi G \rho\,,
\label{Poisson}
\ee
which is not satisfied in the zeroth order approximation, because the potential $\Phi$ is considered as a first order quantity, while the matter density $\rho =\rho_b+\delta \rho $ includes zero (background) and first order terms. This problem is discussed in several textbooks, as e.g. in the aforementioned references~\cite{ZN,Mukha,SW,GR}. It is also noted in an early paper indicated to us by an anonymous referee~\cite{bonnor}, where the author writes: ``In (1.1) the density and pressure are supposed to be uniform throughout the gas. In fact, however, if gravitation is taken into account the equation of hydrodynamic equilibrium has no solution for a finite uniform mass.''

To cure this shortcoming Mukhanov~\cite{Mukha} suggested adding an antigravitating substance, such as e.g. vacuum-like energy, which would counterbalance the gravitational attraction of the background, so that Eq.~(\ref{Poisson}) would be satisfied at zeroth order. Alternatively in Ref.~\cite{Zhuk} the authors assumed that the background density is zero, so that Equation~(\ref{Poisson}) becomes a relation between first order terms. This problem is absent in cosmology, where the zeroth order background equations are satisfied. They are the usual Friedmann equations in an homogeneous, isotropic universe, see for instance Refs.~\cite{ZN,Mukha,SW,GR,bonnor}. 

In this paper we take a different approach to the classical Jeans problem, assuming that Eq.~(\ref{Poisson}) is valid for zeroth order terms
so the solution of the equations of motion leads to time dependent background energy density and gravitational potential. These evolve with time in accordance following the equations of motion. The characteristic timescale of variation of these quantities is close to the Jeans time [both are essentially the gravitational time $t_g \sim (G\rho )^{-1/2} $], so the development of the Jeans instability goes faster than in the standard theory, where such effect is not taken into account. This problem is studied in section~\ref{s-Jeans}.   

The treatment of the Jeans instability in General Relativity starts from the Einstein equations: 
\be
G_{\mu \nu} \equiv R_{\mu \nu} - \frac{1}{2}\, g_{\mu \nu} R = 8\pi G\,T_{\mu \nu} \equiv \tilde T_{\mu \nu} \,.
\label{G-mu-nu}
\ee 
These equations automatically include the equations of motion of matter, namely the continuity and Euler equations. On the other hand the equations of motion of matter can be equivalently obtained from the conditions of covariant conservation of the energy-momentum tensor
\be 
D_{\mu} T^{\mu}_{\nu}=0\, ,
\label{D-T-mu-nu}
\ee  
where $D_{\mu}$ is the covariant derivative in the gravitational field under scrutiny. Usually, it is technically more difficult to derive the Euler and continuity equations from (\ref{G-mu-nu}) because in this case one has to include the terms proportional to the square of the Christoffel symbols in the expression for the Ricci tensor. 

We consider the problem of gravitational instability for an initially spherically symmetric distribution of matter which generates a Schwarzschild-like background gravitational field. In contrast to the usually considered cosmological case, the background metric is not only a function of time but also a function of space coordinates. It leads to difficulties in imposing the Newtonian gauge condition. The problem of gauge fixing and the instability in a coordinate dependent background are studied in section~\ref{s-GR-instab}.  

The final part of the paper, section~\ref{s-F-of-R}, is devoted to gravitational instability in $F(R)$ modified gravity theories. In cosmology this problem was considered in several works for different forms of $F(R)$, see e.g. Refs.~\cite{ref-request,Star-instab,Matsumoto}. 
We thank an anonymous referee for indicating several relevant papers~\cite{ref-request} to us.

An analysis of the Jeans instability for stellar-like objects in modified gravity was performed in Refs.~\cite{Capo,Zhuk}. In these works a perturbative expansion of $F(R)$ was performed either around $R=0$ or $R=R_c$, where $R_c$ is the present cosmological curvature scalar. In our work we expand $F(R)$ around the curvature of the background metric $R_m$, which is typically much larger than $R_c$.

\section{Jeans instability in Newtonian theory with space and time dependent background \label{s-Jeans}}

We consider a spherically symmetric cloud of particles with initially vanishing pressure and velocities, and study the classical non-relativistic Jeans problem in Newtonian gravity. The essential equations are the well known Poisson, Euler, and continuity equations:
\begin{subnumcases}{}
\Delta \Phi = 4\pi G \rho ,
\label{poisson-newt}\\
\partial_t (\rho {\bf v}) + \rho ({\bf v\nabla}) {\bf v} + {\bf \nabla } P  + \rho {\bf  \nabla} \Phi = 0, 
\label{Euler}\\
\partial_t \rho + \nabla (\rho {\bf v}) = 0.
\label{continuity}
\end{subnumcases}

It has been already mentioned in the Introduction that the problem with these equations, as described in the book by Zeldovich and Novikov~\cite{ZN}, is that a time independent $\rho $ is not a solution to these equations. To avoid this problem the authors suggested studying solutions in the cosmological background, while  Mukhanov~\cite{Mukha} proposed to introduce some repulsive force. Instead we consider the time dependent problem taking as initial value a homogeneous distribution $\rho =$ const. inside a sphere with radius $r_m$, while outside this sphere $\rho = 0$. The initial values of particle velocities and pressure are taken to be zero and the potential $\Phi $ is supposed to be a solution of the Poisson equation (\ref{poisson-newt}):
\be
\Phi_0 (r>r_m)  = - MG/r,\,\,\, \Phi_0 (r <r_m) = 2 \pi G \rho_0 r^2 /3 + C_0,
\label{Pho-of-r}
\ee
where $C_0 = - 2\pi G \rho_0 r_m^2$ is chosen such that the potential is continuous (the value of $C_0$ is not important for us), and the total mass of the gravitating sphere is $M = 4\pi \rho_0 r^3_m/3$. 

In what follows we will be interested in the internal solution for $r<r_m$. Now we can find how the background quantities $\rho$, $v$, and $P$ evolve with time at small $t$. From Eq. (\ref{Euler}) it follows that:
\be
v_1 (r,t) = - \nabla \Phi_0 t  = - 4\pi G\rho_0 r t / 3.
\label{V-1}
\ee
From the continuity equation~(\ref{continuity}) we find
\be
\rho_1 = \frac{2\pi}{3} G \rho_0^2 t^2\,\,\, {\rm or}\,\,\, \rho_b (t,r) = \rho_0 + \rho_1 = \rho_0 \left( 1 + \frac{2\pi}{3} G\rho_0 t^2\right).
\label{rho-1}
\ee
It is interesting that $\rho$ rises with time but remains constant in space. Because of the homogeneity of $\rho$ the pressure remains zero, i.e. $P_1 = 0$. 

The time variation of the background potential is found using~(\ref{poisson-newt}):
\be
\Phi_b(r,t) = \Phi_0 + \Phi_1 = \frac{2 \pi}{3} G r^2 \rho_0 \left(1 + \frac{2 \pi}{3} G \rho_0 t^2\right).
\label{phi-1}
\ee

Now we can study the evolution of perturbations over this time-dependent background. We proceed as usual, writing $\rho = \rho_b (r,t) + \delta \rho$, $\Phi = \Phi_b (r,t) + \delta \Phi$, $v = v_1 (r,t) + \delta v$, and $\delta P = c_s^2 \delta \rho$, where $c_s$ is the speed of sound. Here all $\delta$-quantities are infinitesimal and are neglected beyond first order. In what follows we also neglect the products of small sub-one quantities (i.e. $\rho_1$ etc) with $\delta$'s. This significantly simplifies the calculations, while of course the results do not change significantly. We find:
\begin{subnumcases}{}
\Delta (\delta \Phi) = 4\pi G \delta \rho ,
\label{delta-Phi}\\
\partial_t \delta {\bf v} + \nabla \delta \Phi +\delta \rho /\rho_0 \nabla \Phi_b+ \nabla \delta P /\rho_0 = 0,
\label{delta-v} \\ 
\partial_t \delta \rho + \rho_0 \nabla (\delta {\bf v}) = 0.
\label{delta-rho}
\end{subnumcases}

Eq.~(\ref{delta-v}) contains the term $(\delta \rho /\rho_0) \nabla \Phi_b$ which explicitly depends on the coordinate $r$ through the background potential $|\nabla \Phi_b|=4\pi G r \rho_0/3$. We estimate this term substituting instead of $r$ its maximum value $r_m$. To see if this term is essential, let us take the Fourier transform of the last term in Eq.~(\ref{delta-v}): 
\be
\int \frac{d^3 k}{(2\pi )^3} \frac{ \nabla \delta P}{\rho_0} e^{-i\lambda t + i\bf k \bf r} \sim kc_s^2 \, \frac{\delta \rho (\lambda, \bf k)}{\rho _0}\,.
\label{Fourier}
\ee 
So we have to compare $kc_s^2$ with $4\pi G \rho_0/3$.  Evidently, 
\be
4\pi G r_m \rho_0/3 = \frac{r_g}{2r_m^2}\, ,
\ee
where $r_g=2GM$ is the gravitational (Schwarzschild) radius and $M$ is the total mass of the spherical cloud under scrutiny. If $k$ is of the order of the Jeans wave number:
\be
k\sim k_J=\frac{\sqrt{4\pi G\rho_0}}{c_s}\, ,
\label{k-J}
\ee
we can neglect the $r$-dependent term $(\delta \rho /\rho_0) \nabla \Phi_b$ in comparison to $\nabla \delta P /\rho_0$ 
for $c_s>\sqrt{2r_g/(3r_m)}$. There is quite a large volume of the parameter space where this condition is fulfilled. 

Taking the Fourier transform of Eqs.~(\ref{delta-Phi})--(\ref{delta-rho}) and neglecting the $r$-dependent term we obtain the eigenvalue equation:
\be
k^2 (\lambda^2 - c_s^2 k^2 + 4\pi G \rho_0 ) = 0\,.
\label{omega-of-k}
\ee
For small $k$ we find the usual exponential Jeans instability: 
\be
\frac{\delta \rho_{J1}}{\rho_0} \sim \exp \left[ t \left(4\pi G \rho_0 - c_s^2 k^2 \right)^{1/2}\right]. 
\label{delta-rho-J1}
\ee
However, these small perturbations have the same 
characteristic rising time $\sim 1/(4\pi G\rho_0)^{1/2}$ as that of the classical rise of $\rho_1$. 
We can estimate the impact of the rising background energy density on the rise of perturbations making an adiabatic approximation, namely replacing the exponent in Eq.~(\ref{delta-rho-J1}) with the integral: 
\be
\frac{\delta \rho_{J2}}{\rho_0} \sim \exp \left\{   \int^{t}_0 dt \left[4\pi G \rho_b(t, r) - k^2 c_s^2\right]^{1/2}\right\}. 
\label{delta-rho-J2}
\ee   
where $\rho _b (t, r)$ is given by Eq.~(\ref{rho-1}). 

Estimating the above integral for small $k$ we find that the enhancement factor $\delta \rho_{J2}/\delta \rho_{J1}$ is equal to 1.027
after a time $t=t_{grav}$, where $t_{grav}=1/\sqrt{4\pi G \rho_0}$, while for $t=2\,t_{grav}$ it is 1.23, for $t=3\,t_{grav}$ it is 1.89, and for $t=5\,t_{grav}$ it is 11.9. Note that to derive (\ref{delta-rho-J1}) and~(\ref{delta-rho-J2}) we assumed that $t<t_{grav}$, so we should not treat these factors as numerically accurate; still, we can interpret them as an indication that the rise of fluctuations is indeed faster than in the usual Jeans scenario.

\section{Density perturbations in General Relativity \label{s-GR-instab}}

In this section we consider the evolution of scalar perturbations in GR. First we present the necessary expressions for metric, curvature tensors, and energy-momentum tensor of matter, which is taken in the perfect fluid form. Next we discuss the choice of gauge for perturbations in the coordinate-dependent background. And lastly we study the rise of perturbations in a Schwarzschild-like background, which may depend on both time and space coordinates.      

\subsection{Metric and Curvature \label{ss-metric}}

As we did previously, we consider a spherically symmetric cloud of matter with initially homogeneous energy density inside the limit radius $r=r_m$. We choose Schwarzschild-like isotropic coordinates in which the metric takes the form:
\be
ds^2=A\,dt^2 - B\,\delta_{ij}\,dx^i dx^j\,, 
\label{ds-2}
\ee
where the functions $A$ and $B$ may depend upon $r$ and $t$. The corresponding Christoffel symbols are:
\be 
&&\Gamma ^t_{tt}=\frac{\dot A}{2A}\,,\ \ \ \Gamma ^t_{jt}=\frac{\partial_j A}{2A}\,, \ \ \ 
\Gamma ^j_{tt}=\frac{ \delta^{jk} \partial_k A}{2B}\,, \ \ \ 
\Gamma ^t_{jk}=\frac{ \delta_{jk} \dot B}{2A}\,, \nonumber \\
&&\Gamma ^k_{jt}=\frac{ \delta^k_j \dot B}{2B}\,, \ \ 
\Gamma ^k_{lj}=\frac{1}{2B}(\delta^k_l\partial_j B + \delta^k_j\partial_l B - \delta_{lj}\delta^{kn}\partial_n B)\,.  
\label{Gammas}
\ee

For the Ricci tensor, including terms quadratic in $\Gamma$'s, we obtain:
\be
R_{tt} &=& \frac{\Delta A}{2B} - \frac{3 \ddot B}{2B} + \frac{3 \dot B^2}{4B^2} + \frac{3 \dot A \dot B}{4AB} +
\frac{\partial^jA \partial_j B}{4B^2} - \frac{\partial^jA \partial_j A}{4AB}\, , \\
\label{R-tt}
R_{tj} &=& - \frac{\partial_j \dot B}{B} + \frac{\dot B \partial_jB}{B^2} + \frac{\dot B \partial_jA}{2AB}\, , \\
\label{R-tj}
R_{ij} &=& \delta_{ij} \left(\frac{\ddot B}{2A} - \frac{\Delta B}{2B} + \frac{\dot B^2}{4AB} -\frac{\dot A \dot B}{4A^2}
- \frac{\partial^kA \partial _kB}{4AB} +   \frac{\partial^kB \partial _kB}{4B^2}   \right)   \nonumber \\
&-&\frac{\partial_i\partial_j A}{2A} - \frac{\partial_i\partial_j B}{2B} +
\frac{\partial_i A\partial_j A}{4A^2} + \frac{3\partial_i B\partial_j B}{4B^2} +
\frac{\partial_i A\partial_j B + \partial_j A\partial_i B}{4AB}\, .
\label{R-ij}  
\ee  
Here and in what follows the upper space indices are raised with the Kronecker delta, $\partial^j A = \delta^{jk} \partial_k A$. 

The corresponding curvature scalar is:
\be 
R= \frac{\Delta A}{AB} - \frac{3 \ddot B}{AB} + \frac{2\Delta B}{B^2}+
\frac{3\dot A\dot B}{2A^2B} - \frac{\partial^jA\partial_jA}{2A^2B} - \frac{3\partial^jB\partial_jB}{2B^3} +
\frac{\partial^jA\partial_jB}{2AB^2}\,.
\label{R}
\ee   
Let us now present the expressions for the Einstein tensor $G_{\mu\nu}=R_{\mu \nu} - 1/2 \,g_{\mu \nu} R$:
\be 
G_{tt}&=&-\frac{A\Delta B}{B^2}+\frac{3\dot B^2}{4B^2}+\frac{3A \partial^jB\partial_jB}{4B^3}\,, \label{G-tt} \\
G_{tj}&=&R_{tj}\,, \label{G-tj} \\
G_{ij}&=&\delta_{ij} \left(\frac{\Delta A}{2A}+\frac{\Delta B}{2B} - \frac{\ddot B}{A} + \frac{\dot B^2}{4AB} +
\frac{\dot A \dot B}{2A^2}
- \frac{\partial^kA \partial _kA}{4A^2} -  \frac{\partial^kB \partial _kB}{2B^2}   \right)   \nonumber \\
&-&\frac{\partial_i\partial_j A}{2A} - \frac{\partial_i\partial_j B}{2B} +
\frac{\partial_i A\partial_j A}{4A^2} + \frac{3\partial_i B\partial_j B}{4B^2} +
\frac{\partial_i A\partial_j B + \partial_j A\partial_i B}{4AB}\, .
\label{G-ij}
\ee


The energy-momentum tensor is taken in the perfect fluid form without dissipative corrections:
\be 
T_{\mu\nu}=(\rho + P)U_{\mu}U_{\nu} - P g_{\mu \nu}\,,
\label{T-mu-nu}
\ee
where $\rho $ and $P$ are respectively the energy density and pressure of the fluid and the four-velocity is:
\be
U^{\mu}=\frac{dx^{\mu}}{ds} \ \ \ {\rm and} \ \ \ U_{\mu}=g_{\mu \alpha}U^{\alpha}\,.
\label{U-mu}
\ee
We assume that the three-velocity $v^j = dx^j/dt$ is small and thus neglect terms quadratic in $v$. Correspondingly, 
\be
U_j = - \frac{B v_j}{\sqrt{A} \sqrt{1-(B/A) v_j v^j}} \approx  - \frac{B v_j}{\sqrt{A}}\,.
\label{U-j}
\ee
According to our definition $v_j=v^j$. From the condition 
\be
1=g^{\mu\nu} U_{\mu}U_{\nu} = \frac{1}{A}U_t^2 - \frac{1}{B}\delta^{kj} U_k U_j \approx  \frac{1}{A}U_t^2
\ee
we find $U_t \approx 1/\sqrt{A}$. Now we can write:
\be
T_{tt} &=& (\rho +P) U_t^2 - P A \approx \rho A\,, \\
\label{T-tt}
T_{jt}  &=& (\rho +P) U_t U_j \approx - (\rho +P) B v_j\,, \\
\label{T-jt}
T_{ij} &=&  (\rho +P) U_i U_j - P g_{ij} \approx P B \delta_{ij}\,.
\label{T-ij}
\ee 

\subsection{Choice of gauge \label{ss-gauge}}

In a cosmological situation, the spatially flat Friedmann background  depends only on time and not on the space coordinates: 
\be
ds^2_{cosm}= dt^2 - a^2(t) d{\bf r^2} =  a^2(\eta )(d\eta^2 - d{\bf r}\,^2)\,,  
\label{ds2-cosm}
\ee
As shown in several textbooks, see e.g.~\cite{Mukha,SW,GR}, this allows to impose the Newtonian gauge condition on the perturbed metric, which for scalar perturbations takes the form:
\be
ds^2_{pert}= (1 + 2\Phi) dt^2 - a^2(t) (1 - 2 \Psi)\,\delta_{ij}\,dx^i dx^j\,,
\label{ds2-pert}
\ee
where $\Phi$ and $\Psi$ are the metric perturbations or, in other words, stochastic deviations from the cosmological background.

In our case the background metric is taken as the internal Schwarzschild metric in isotropic coordinates, see e.g. Ref.~\cite{Sch-int}, chapter 16:
\be
ds^2_{Sch}= A\,dt^2 -  B\,\delta _{ij}\,dx^idx^j\, ,
\label{ds2-Sch}
\ee
where $ A$ and $ B$ are functions of space and time in the form
\be
A(t,r)=1+a(t)r^2\,,\qquad B(t,r) = 1+b(t)r^2\,.
\ee
Calculations will be greatly simplified assuming that deviations from a flat Minkowski metric are sufficiently
weak and so $ A\approx 1$ and $B\approx 1$. The dependence of the background on space coordinates generates serious problems when one
tries to impose the Newtonian gauge condition, as we illustrate in what follows.

For scalar fluctuations the general form of the perturbed metric is:
\be 
ds^2_{scalar} = (A + 2\Phi) dt^2 + (\partial _j C) dt\,dx^j - \left[(B -2\Psi )\delta _{ij} - \partial _i  \partial _j E\right] dx^i dx^j \,.
\label{ds-scalar} 
\ee
The Newtonian gauge condition implies $C=E=0$, which can be easily realised in cosmology by a proper change of coordinates. Under the coordinate transformation $\tilde x^{\alpha } = x^{\alpha } + \xi ^{\alpha }$ the metric tensor transforms as
\be
\tilde g_{\alpha \beta} (\tilde x) =  g_{\alpha \beta}^b (\tilde x)+ \delta g_{\alpha \beta} - g_{\alpha \mu}^b \partial_\beta \xi ^{\mu} - g_{\beta \mu}^b \partial_\alpha \xi ^{\mu}\,,
\label{tilde-g-alpha-beta}
\ee  
where $g_{\alpha\beta}^b$ is the ``old'' background metric at the point $\tilde x$ and the $\delta g_{\alpha \beta} $ are the fluctuations around this metric. Fluctuations around the new metric are defined as $\delta \tilde g_{\alpha \beta} = \tilde g_{\alpha \beta}(\tilde x) - g_{\alpha \beta}^b(\tilde x)$. Taking into account that $g_{\alpha \beta}^b(\tilde x) = g_{\alpha \beta}^b( x) + (\partial _{\mu} g_{\alpha \beta} ^b )\xi ^{\mu}$, we finally find:
\be
\delta \tilde g_{\alpha \beta}  =  \delta g_{\alpha \beta} - (\partial _{\mu} g_{\alpha \beta} ^b) \xi ^{\mu}
 - g_{\alpha \mu}^b \partial_\beta \xi ^{\mu} - g_{\beta \mu}^b \partial_\alpha \xi ^{\mu}\,.
 \label{delta-tilde-g}
\ee 
This gives:
\be
\delta \tilde g_{00} &=& \delta g_{00} - (\xi^t \partial_t A + \xi^k \partial_k A) - 2A \partial _t \xi ^t \,, 
\label{delta-00} \\ 
\delta \tilde g_{0j} &=& \delta g_{0j} - A \partial_j\xi^t + B\delta_{jk}\partial_t \xi^k\,, 
\label{delta-0j} \\
\delta \tilde g_{ij} &=& \delta g_{ij} + \delta_{ij} (\xi^t \partial_t B + \xi^k \partial_k B) + B(\delta_{kj} \partial_i \xi^k + \delta_{ki} \partial_j \xi^k)\,.
\label{delta-ij}
\ee
For scalar perturbations we restrict ourselves to ``longitudinal'' coordinate changes, that is: 
\be
\xi^ i = \partial^i \zeta = - \partial_j \zeta/B\,.
\label{xi-j-long}
\ee
To eliminate $\delta \tilde g_{0j}$ we have to impose the condition:
\be
\partial_jC - A\,\partial_j \xi^t - B\,\partial_t (\partial_j \zeta/ B) \equiv  
 \partial_j \left[C - B\,\partial_t ( \zeta/B) - A\,\xi^t \right] + \partial_j B\, \partial_t (\zeta/B) +
\xi^t \partial_jA = 0\,.
\label{delta-0j=0}
\ee
The sum of the last two terms in this equation vanishes if we choose $\xi^t = (B'/A')\, \partial_t (\zeta/B) $, where a prime denotes derivative with respect to $r$. Evidently the term in square brackets can be cancelled out with a proper choice of $\zeta $.

Now we need to get rid of the gradient terms in Eq. (\ref{ds-scalar}), i.e. to impose the condition:
\be
\partial _i \partial_j E - 2 \partial _i \partial _j \zeta + \frac{\partial _i B}{B}\partial _j \zeta + \frac{\partial _j B}{B}\partial _i \zeta 
 = 0\,.
\label{delta-ij=0}
\ee   
Unfortunately, there is no way to satisfy this equation. Firstly, we have already used all the freedom to eliminate $g_{tj}$ and, secondly,   
there are terms of two different kinds. The first two terms are purely longitudinal ones, while the last two contain both transverse and
longitudinal contributions and it is impossible to eliminate both with a single function $\zeta$.

Let us note that with the ``scalar'' coordinate change there appear vector and tensor metric perturbations due to the dependence of the background
metric functions on the spatial coordinates. This is an artefact of the coordinate choice.  Probably these vector and tensor modes could be eliminated if one allows for a ``transverse'' coordinate transformation $\xi^j=  \xi^j_\perp + \partial^j \zeta$. We will not pursue this issue further and in what follows we will assume, as we 
have mentioned above, that deviations from the flat metric are small and thus $A\approx B \approx 1$. In this approximation the problems with the gauge do not appear. 
The matter of the gauge choice in the case of the coordinate-dependent background will be studied elsewhere. Other possible gauges used for the study of gravitational instability and the conditions of their validity are described in Ref.~\cite{much-phrept} (see also the book~\cite{Mukha}). It is possible that other gauge choices may be more suitable to study the space dependent case.

\subsection{Evolution of fluctuations in General Relativity \label{ss-GR-instab}}

Usually the GR equations are taken in the weak field limit, so the terms proportional to $\Gamma^2$ in the expressions for the Ricci tensor are neglected. Differentiating the GR equation for $G_{tt}$ over time and the one for $G_{jt}$ over $x^j$ we derive the continuity equation, while taking the time derivative of the equation for $G_{jt}$ and the derivative over $x^i$ of the equation for $G_{ij}$ we obtain the Euler equation. However, if we restrict ourselves to the first order in $\Gamma$ in the Ricci tensor we do not obtain self-consistent equations. So, the second order terms in $R_{\mu\nu}$ are necessary and we derived the continuity and Euler equations by this procedure. 

On the other hand, one can take a simpler path, deriving the Euler and continuity equations from the conditions $D_{\mu} T^{\mu}_j=0$ and $D_{\mu} T^{\mu}_t=0$. Since we have four unknown functions we need two more equations as which we can take the equation for $G_{tt}$ and the $\partial _i \partial _j$-component of the equation for $G_{ij}$, at linear order in $\Gamma$'s. Correspondingly we keep only terms linear in the derivatives of $A$ and $B$ and take $A=B=1$ otherwise. 

The equations for $G_{tt}$ and for the $\partial _i \partial _j$-component of the equation for $G_{ij}$ are:
\begin{subequations}
\begin{align}
-\Delta B = \tilde \rho\,,
\label{tt-1}\\
\partial _i \partial _j(A+B)=0\,.
\label{ij-1}
\end{align}
The continuity and Euler equations are respectively:
\begin{align}
&\dot \rho  + \partial_j[(\rho + P)v^j] + \frac{3}{2}\rho\dot B=0\,,
\label{dot-rho}\\
&\rho\,\dot v_j + \partial_j P + \frac{1}{2} \rho\, \partial_j A  =0\,.
\label{dot-v}
\end{align}
\end{subequations}

We assume that the background metric changes slowly as a function of space and time and study small fluctuations around the background quantities: $\rho = \rho_b + \delta \rho$, $\delta P =c_s^2 \delta \rho$, ${\bf v} = \delta {\bf v}$, $A=A_b+\delta A$, $B=B_b+\delta B$.   

The corresponding linear equations for the infinitesimal perturbations are:
\begin{subnumcases}{}
-\Delta \delta B = \delta \tilde \rho\,,
\label{delta-tt-1}\\
\partial _i \partial _j(\delta A+\delta B)=0\,. 
\label{delta-ij-1}\\
\delta\dot \rho  + \rho\, \partial_j \delta v^j + \frac{3}{2}\rho\,\delta \dot B=0\,,
\label{delta-dot-rho}\\
\rho\, \delta \dot v_j + \partial_j \delta P + \frac{1}{2}\rho \,\partial_j \delta A  =0\,.
\label{delta-dot-v}
\end{subnumcases}
Eqs.~(\ref{delta-tt-1}) - (\ref{delta-dot-v}) coincide with the corresponding equations in books \cite{SW,Mukha,GR} for a static universe, i.e. for $a(t)=1$ and $H=0$. Remember that with our definitions: $\delta A \equiv 2\Phi$ and $\delta B \equiv - 2\Psi $.

We look for solution in the form $\sim \exp[-i \lambda t + i\bf k \cdot \bf x]$ and obtain the following expressions for the frequency eigenvalues:
\be
\lambda^2 = \frac{c_s^2 k^2 - \tilde \rho /2}{1 + 3 \tilde \rho/ (2k^2)}\,.
\label{lambda-GR}
\ee 
This result almost coincides with the Newtonian one~(\ref{delta-rho}). An extra term in the denominator is induced by the volume variation, and it is small for $k \sim k_J$ [see Eq.~(\ref{k-J})].

\section{Instability in modified gravity \label{s-F-of-R}}

Gravitational instability in the framework of modified gravity was studied in detail in Ref.~\cite{Capo} for the case of star formation with realistic pressure. We consider the simpler situation of pressureless gas which occurs at an initial stage of galaxy or star formation. We did not impose the condition of $F(R=0)=0$ used in the cited works, or expand the theory around such value, because we are interested in higher density configurations (see below).

In modified gravity a nonlinear function of curvature $F(R)$ is added to the standard Einstein-Hilbert Lagrangian, so the new Einstein equations have the form: 
\be
\left( 1 + F_{,R}\right) R_{\mu\nu} -\frac{1}{2}\left( R + F\right)g_{\mu\nu}
+ \left( g_{\mu\nu} D_\alpha D^\alpha - D_\mu D_\nu \right) F_{,R}  = 
\frac{8\pi T_{\mu\nu}}{m_{Pl}^2} \equiv \tilde T_{\mu\nu}\,,
\label{eq-of-mot}
\ee 
where $F_{,R} = dF/dR$. Some examples of $F(R)$-theories which realistically describe the cosmological acceleration can be 
found in Refs.~\cite{Starob,HuSaw,ApplBatt,App-Bat-Star,F-of-R-rev,F-of-R-rev_Noj_Odin}.

In these models, the function $F(R)$ takes very different values for $|R|\ll |R_c|$, $|R|\sim |R_c|$, and $|R| \gg |R_c|$, where $R_c$ is the present average cosmological curvature. We consider the case $|R| \gg |R_c|$, which is realised in astronomical systems with the energy density grossly exceeding the cosmological one. In many models presented in the literature the following conditions are also fulfilled in this regime: $|F| \ll |R|$ and $|F_{,R}| \ll 1$. In this case the equations of motion are greatly simplified and instead of Eqs.~(\ref{delta-tt-1}) and (\ref{delta-ij-1}) we arrive to:
\begin{subequations}
\be
&&-\Delta \delta B + \omega^{-2} \Delta \delta R= \delta \tilde \rho\,,
\label{delta-tt-2}\\
&&\partial _i \partial _j(\delta A+\delta B - 2\omega^{-2} \delta R)=0\,,
\label{delta-ij-2}
\ee
\end{subequations}
where
\be
\omega^2\equiv -\frac{1}{3F_{,RR}}
\ee
is the typical frequency associated with the $F(R)$ model under scrutiny. Equations (\ref{delta-dot-rho}) and (\ref{delta-dot-v}) remain unmodified, thus it is straightforward to use the results of sec.~\ref{s-Jeans} about perturbations in time dependent background.

Taking the usual Fourier transform $\sim \exp[-i \lambda t + i\bf k \cdot \bf x]$, we obtain the following four equations for the four unknowns $\delta A$, $\delta B$, $\delta \tilde \rho$, and the longitudinal component of velocity $\delta v_j$:
\begin{subnumcases}{}
\delta A + \delta B = 2 \omega^{-2}\delta R \,, \\
k^2 (\delta B - \delta A) = 2 \delta \tilde \rho \,, \\
k_j \lambda\,\tilde \rho \, \delta v_j - k^2 c_s^2 \delta \tilde \rho - k^2 \tilde \rho \, \delta A/2=0 \,, \\
k_j \lambda\,\tilde \rho \, \delta v_j -\lambda^2 \delta \tilde \rho -3 \lambda^2\tilde \rho \, \delta B/2 = 0\,,
\end{subnumcases} 
where [see Eq.~(\ref{R})]
\be
\delta R=(3\lambda^2 - 2k^2) \delta B -k^2 \delta A\,.
\ee

Eliminating the term $k_j \lambda\,\tilde \rho\,\delta v_j $ from the two last equations we arrive at the following equation for $\lambda$:
\be
\frac{3k^2}{\omega^2}\,\lambda^4 - \lambda ^2 \left[k^2 + \frac{3\tilde \rho}{2} + 
\frac{3k^4}{\omega^2}\left(1 + c_s^2\right)\right] - \frac{\tilde \rho k^2}{2} + c_s^2 k^4  + 
\frac{k^4}{\omega^2}\left(3c_s^2 k^2 - 2\tilde \rho\right)=0\,.
\label{lambda-of-k}
\ee 
The solution of Eq.~(\ref{lambda-of-k}) is straightforward but very tedious because of the many relevant parameters. The sign of the
product of the roots is determined by the last free term in this equation: 
\be
\frac{\lambda_1^2 \lambda_2^2}{c_s^2\omega^4} = \kappa^4 + \frac{\kappa^2(1-\gamma)}{3} -\frac{\gamma}{12}\,,
\label{prod-lambda}
\ee
where $\kappa = k/\omega$ and $\gamma = 2 \tilde \rho/ (\omega^2 c_s^2)$. It is interesting that the product of the
eigenfrequencies  depends on a single parameter $\gamma$.

In the limit  $k^2 \ll \omega^2$ we obtain the following simple expressions for the eigenvalues:
\be
\lambda_1^2 &=& \frac{c_s^2 k^2 - \tilde \rho /2}{1 + 3 \tilde \rho/ (2k^2)}\,, 
\label{lambda-1-MGr} \\
\lambda_2^2 &=& \omega^2 \left(\frac{1}{3} + \frac{\tilde \rho}{2k^2}\right)\,.
\label{lambda-2-MGr}
\ee 
The first root coincides with the usual result of GR, Eq.~(\ref{lambda-GR}), while the second root is similar to the high frequency solutions 
found in modified gravity by a different approach~\cite{ADR-1,ADR-2}. In the present day astrophysical or cosmological background the
characteristic frequency of oscillations is much lower than the scalaron mass in the early universe. For example, for $F(R)$ of the 
models~\cite{Starob,HuSaw,ApplBatt} the frequency is $\omega \sim (1/t_U) (R/R_c)^{n+1} $, which is typically much smaller than $m$, 
though it may approach $m$ with rising $n$. Anyhow the particle production rate is much smaller than the characteristic frequency because it is suppressed by a power of the Planck mass, $\Gamma \sim \omega^3/m_{Pl}^2$, see e.g. Refs.~\cite{ADR-1,ADR-2,part-prod}. So the
oscillations are not effectively damped.

We can take into account possible time variations of the background treating them adiabatically in the spirit of Sec.~\ref{s-Jeans}, see Eq.~(\ref{delta-rho-J2}). This approach gives a reasonable estimate for the speed of variation of $\delta \rho/\rho$ if the background is slowly changing. So we can substitute $\rho_b = \rho_b(t)$. However, for fast variations of the background quantities, as for the high frequency curvature oscillations found in our works~\cite{ADR-1,ADR-2}, the adiabatic approximation does not work and the solution for $\delta\rho$ is strongly modified. We are currently investigating this problem.

It is instructive to present the expression for the Jeans wave number with the correction induced by modified gravity. As can be seen from
Eqs.~(\ref{lambda-1-MGr}, \ref{lambda-2-MGr}) in the limit of low wave number $k \ll \omega^2$ the usual results of GR are recovered.
But there is an additional eigenfrequency, which does not disappear even if the correction to GR is made arbitrary weak. This is related to the
higher order of the equations of motion. These modes are purely oscillatory and stable. 

There are some corrections to the Jeans wave vector, $k_J$, if $\omega$ is finite. Since $k_J$ corresponds to the vanishing eigenfrequency
$\lambda =0$, it follows from Eq.~(\ref{prod-lambda}) that the dimensionless quantity $\kappa_J = k_J/\omega$ satisfies the equation:
\be
\kappa^4_J + \frac{\kappa^2_J(1-\gamma)}{3} - \frac{\gamma}{12} = 0\,,
\label{eq-for-kappa}
\ee
Eq.~(\ref{eq-for-kappa}) is easy to solve and one can see that the gravity modification leads to an increase of $k_J$. In particular for
large $\omega$ or $\gamma \ll 1$ we obtain
\be
k_J^2 = \frac{\tilde \rho}{2 c_s^2} \left( 1 + \frac{\tilde \rho}{2 \omega^2 c_s^2} \right).
\label{k-J-1}
\ee
In the opposite limit, $\gamma \gg 1$, we find $ k_J^2 = \omega^2 (\gamma  - 1/4)/3$.

The other solution of Eq. (\ref{eq-for-kappa}) gives $k^2_J < 0$. It describes fluctuations varying with coordinates, similar to waves propagating in wave-guides. Such solution may possibly be physically realised in confined systems, as e.g. collapsing stars. The frequency eigenvalue $\lambda$ might be imaginary for certain negative values of $k^2$, which would lead to a new type of instability. This is an interesting case and is deserving of further study.

\section{Conclusions}
We have found that taking into account the increase of the background energy density in the classical Jeans problem leads to a faster growth of perturbations. The obtained results are valid for a cloud at an initial stage of galaxy or star formation when the background pressure can be neglected.

The choice to the Newtonian gauge for scalar perturbations in cosmology when the metric functions depend only on time is easily achieved. However, if the metric functions depend upon space coordinates, one encounters serious technical problems. In particular, vector and tensor modes may be induced as a result of the coordinate transformation. The GR result for the evolution of perturbations essentially coincides with that of the classical Jeans theory with a small correction accounting for a change of volume due to the time variation of the background metric.

The evolution of the density perturbations in modified gravity contains a new high frequency mode because the equations of motion in this case are higher order. The existence of such mode was anticipated in our works \cite{ADR-1,ADR-2}. In Ref.~\cite{ADR-anti} it was found that in the background of quickly oscillating solutions, gravitational repulsion in finite size objects is possible. A study of the Jeans-like instability over such quickly oscillating background is in progress.

\acknowledgments
EA and AD acknowledge the support of the grant of the Russian Federation government 11.G34.31.0047. EA thanks organisers and participants of the workshop SW8--``Hot topics in modern cosmology'' for stimulating discussions.

\end{document}